\documentclass[journal,twocolumn]{IEEEtran}
\usepackage{epsfig,makeidx,color}
\usepackage{amsmath,amssymb,bbm}
\usepackage{cite,graphicx,lipsum}
\usepackage{enumerate}
\usepackage[switch,pagewise]{lineno}
\usepackage{hyperref}
\hypersetup{
        colorlinks = true,
        citecolor=blue,
}
\newcommand{\indicator}{\mathbbm{1}}

\def\cK{{\cal K}}

\def\rH{{\rm H}}
\def\rT{{\rm T}}

\def\uZ{{\mathbb Z}}
\def\uP{{\mathbb P}}

\def\uC{{\mathbb C}}
\def\uE{{\mathbb E}}

\DeclareMathOperator*{\argmax}{\arg\!\max}

\newtheorem{mylemma}{\bf Lemma} 

\def\deft{ \buildrel \triangle \over = }

\def\be{ \begin{equation} }
\def\ee{ \end{equation} }
\def\bea{ \begin{eqnarray} }
\def\eea{ \end{eqnarray} }

\def\bx{{\bf x}}

\def\bc{{\bf c}}

\def\bb{{\bf b}}

\def\bg{{\bf g}}

\def\ba{{\bf a}}

\def\bn{{\bf n}}
\def\bh{{\bf h}}

\def\bv{{\bf v}}

\def\bI{{\bf I}}

\def\bN{{\bf N}}

\def\bR{{\bf R}}

\def\bY{{\bf Y}}
\def\bZ{{\bf Z}}

\def\b0{{\bf 0}}

\def\cC{{\cal C}}

\def\cN{{\cal N}}

\def\sSINR{{\sf SINR}}

\ifCLASSOPTIONonecolumn
  \newcommand{\figwidth}{0.50\columnwidth}
  
\else
  \newcommand{\figwidth}{0.90\columnwidth}
  
\fi

\begin{document}

\title{On Throughput Improvement using Immediate Re-transmission in 
Grant-Free Random Access with Massive MIMO}

\author{Jinho Choi\\
\thanks{The author is with
the School of Information Technology,
Deakin University, Geelong, VIC 3220, Australia
(e-mail: jinho.choi@deakin.edu.au).
This research was supported
by the Australian Government through the Australian Research
Council's Discovery Projects funding scheme (DP200100391).}}


\maketitle
\begin{abstract}
To support machine-type communication (MTC),
massive multiple-input multiple-output (MIMO) 
has been considered for grant-free random access.
In general, the performance of grant-free random access
with massive MIMO is limited by the number of preambles 
and the number of active devices.
In particular, when there are a number of active devices
transmitting data packets simultaneously,
the signal-to-interference-plus-noise ratio (SINR)
cannot be high enough for successful decoding.
In this paper, in order to improve performance,
we consider immediate re-transmissions for an active device that 
has a low SINR although it
does not
experience preamble collision to exploit
re-transmission diversity (RTD) gain.
To see the performance of the proposed approach,
we perform throughput analysis with certain approximations
and assumption. Since the proposed approach can be unstable due to immediate
re-transmissions, conditions for stable systems are also studied.
Simulations are carried out and 
it is shown that analysis results reasonably
match simulation results.
\end{abstract}

\begin{IEEEkeywords}
Grant-free Random Access; Massive MIMO
Re-transmission; Throughput Analysis 
\end{IEEEkeywords}

\ifCLASSOPTIONonecolumn
\baselineskip 30pt
\fi

\section{Introduction}

In order to provide devices' connectivity for the Internet of Things (IoT), 
a number of different approaches are considered \cite{Ding_20Access}.
Among those, for wide area coverage,
machine-type communication (MTC) in 
5th generation (5G) cellular systems
has been actively studied \cite{Bockelmann16} \cite{Dawy17}.
In MTC, although there are a large number of devices to be connected,
in general, they have short data packets with sporadic activity.
Thus, random access 
based on ALOHA \cite{Abramson70}
is employed for MTC thanks to low signaling overhead
\cite{3GPP_MTC} \cite{3GPP_NBIoT}
\cite{Chang15} \cite{Choi16}.

To support massive connectivity for
a large number of devices and sensors within a cellular system
(e.g., 5G),
massive multiple-input multiple-output (MIMO)
\cite{Marzetta10} can be considered
with a base station (BS) equipped with a large number of antenna
elements. According to
\cite{Bjornson18}, massive MIMO would be a promising solution to
massive MTC, as the capacity becomes unbounded in the presence
of pilot contamination, which can support a very large
number of devices in each cell.
In massive MIMO, the channel state information (CSI) of
each device is characterized by a random channel vector.
Thus, as long as the channel vectors of active devices,
which are the devices with data packets to transmit,
are different (and nearly orthogonal with a sufficiently
large number of antennas), the BS is able to 
detect/decode simultaneously transmitted signals from them.
Thus, for random access with massive MIMO,
it is crucial to allow the BS to estimate the
active devices' channel vectors. To this end,
an active device needs to transmit a preamble
prior to transmission of data packet. Furthermore,
no handshaking protocol is required to
allocate channel resources for transmissions of data packets
from active devices thanks to
high spatial resolution or selectivity in massive MIMO.
As a result, various grant-free
random access schemes with massive MIMO
are studied in \cite{deC17}  \cite{Senel18} \cite{Liu18}
\cite{Ding19_IoT} \cite{Ding21a}.


In most grant-free random access schemes with massive MIMO have two phases. 
In the first phase, each active device
sends a preamble that is randomly chosen from a predetermined
preamble pool (i.e., a finite set of sequences) so that 
the BS is able to estimate active devices' channel vectors.
In the second phase,
the active devices transmit their data packets and
the BS performs beamforming
with estimated channel vectors to decode data packets
from multiple active devices.

In grant-free random access with massive MIMO,
as mentioned earlier, each active device transmits
a preamble. Thus, the BS needs to 
detect transmitted preambles, which is
called device activity detection \cite{Senel18} \cite{Liu18}.
In addition, as discussed in \cite{Jiang19} \cite{Ding20b} \cite{Choi20a},
preamble design plays a key role
in providing a good performance in terms of the probability of successful
decoding or success probability.

Provided that 
the BS is able to detect transmitted preambles without errors
and a set of orthogonal preambles are used,
the performance of 
grant-free random access with massive MIMO
is mainly dependent on 
the number of preambles and the 
number of active devices.
Since the number of preambles is finite,
there is a non-zero probability that 
multiple active devices choose the same preamble, which leads
to preamble collision (PC).
Since the estimated channel vector is a
noisy superposition of multiple channels vectors
of the active devices that choose the same preamble, 
the signal-to-interference-plus-noise ratio (SINR)
after beamforming becomes poor and subsequent decoding for data packet
is likely unsuccessful.

As discussed in \cite{Ding21a},
with massive MIMO, the error probability
of preamble detection is negligible.
Thus, the BS can broadcast
the outcomes of preamble detection
so that active devices with PC do not transmit their data packets,
which can result in a high 
SINR of the other active devices without PC and an improved throughput.
However, if the number of the active devices without PC
is large, the resulting SINR can be low.
Thus, the performance is also limited by the number of active devices
(even if the number of preambles is sufficiently large
for a low probability of PC).

In this paper, we propose a 
grant-free random access scheme with massive MIMO,
where an active device that does not experience PC, but
has a low SINR performs immediate re-transmissions.
Since the active device's channel vector is successfully
estimated by the BS (as PC does not happen),
this device can re-transmit data packets without 
sending additional preambles. This approach
is advantageous for forthcoming active devices, 
since the probability of the PC in subsequent slots 
is not increased by this active device.
In addition, since multiple copies of a data packet
are received by immediate re-transmissions,
the BS can have re-transmission diversity (RTD) gain that
can increase the SINR.
As a result, after a certain number of 
immediate re-transmissions, the BS can eventually
decode the data packet.

Due to immediate re-transmissions,
the proposed system can be unstable
if the number of re-transmissions is unbounded.
Thus, we study the stability issue 
of the proposed system with the throughput analysis.
Unfortunately, there are a number of difficulties.
In particular, since the SINR is a random variable that depends on
the number of active devices (although the channel hardening
effect of massive MIMO is taken into account
\cite{Marzetta10} \cite{Chen18})
and increased by RTD (via immediate re-transmissions),
the throughput analysis is not straightforward.
Thus, for tractable analysis,
we consider approximations of SINR with immediate re-transmissions
and study Foster-Lyapunov criteria
\cite{Kelly_Yudovina} \cite{HajekBook},
which allow us to derive key conditions for stable systems.
Simulations are carried out and it is shown that
the analysis results reasonably agree with simulation results.

In summary, the key contributions of the paper are
\emph{i)} a grant-free random access scheme with massive
MIMO is proposed using immediate re-transmissions to improve 
the throughput;
\emph{ii)} its throughput 
analysis is carried out using Foster-Lyapunov criteria
to find key conditions for stable systems and throughput
under certain conditions.

The rest of the paper is organized
as follows. In Section~\ref{S:SM},
we present the system model for
grant-free random access with massive MIMO.
In Section~\ref{S:SINR}, a simple expression for the SINR 
with estimated channel vector is derived to be used
for throughput analysis.
We discuss different types of feedback signals in 
grant-free random access with massive MIMO and
propose a modified approach that can improve the 
throughput using immediate re-transmissions
in Section~\ref{S:IRT}.
In Section~\ref{S:Anal}, the throughput is analyzed 
using certain approximations and assumptions
and the stability issue of the proposed approach
is addressed.
In Section~\ref{S:Sim}, we present simulation results
that are also compared with the analysis results
obtained in Section~\ref{S:Anal}.
The paper is concluded in Section~\ref{S:Con} with some remarks.

\subsubsection*{Notation}
Matrices and vectors are denoted by upper- and lower-case
boldface letters, respectively.
The superscripts $\rT$ and $\rH$
denote the transpose and complex conjugate, respectively.
$\lfloor x \rfloor$ 
represents the greatest 
integer that is less than or equal to $x$,
and $\lceil x \rceil$ denote the least integer that 
is greater than or equal to $x$.
$\uE[\cdot]$
and ${\rm Var}(\cdot)$
denote the statistical expectation and variance, respectively.
$\cC \cN(\ba, \bR)$
represents the distribution of
circularly symmetric complex Gaussian (CSCG)
random vectors with mean vector $\ba$ and
covariance matrix $\bR$.

\section{System Model}	\label{S:SM}

Suppose that a grant-free random access system consists of
a BS and multiple devices.
It is assumed that the BS is equipped with
$M$ antenna elements and each device has a single antenna.

For grant-free random access,
we consider two phases \cite{Ding19_IoT} \cite{Jiang19}. 
The first phase is the preamble
transmission phase and the following one is the data
transmission phase.  Throughout the paper,
it is assumed that a slot is divided
into two sub-slots, where preambles
are transmitted in the first sub-slot and data packets
are transmitted in the second sub-slot.
In the preamble transmission phase (during the first sub-slot),
each active device with data is to send
a randomly selected preamble from the following preamble pool:
\be
\cC = \{\bc_1, \ldots, \bc_L\},
\ee
where $\bc_l \in \uC^{N \times 1}$ represents the $l$th preamble. 
Throughout the paper, 
it is assumed that the $\bc_l$'s are orthonormal sequences
of length $N$.
Thus, we assume that $L = N$.
After sending a preamble, an active device
sends its data packet in
the data transmission phase.
Throughout the paper, it is assumed that the 
lengths of data packets of all devices
are the same. In addition, all the devices
are synchronized (to this end, the BS needs periodically broadcast
a beacon signal).

Let $\bh_k \in \uC^{M \times 1}$ represent the channel vector 
from active device $k$ to the BS.
Then, in the preamble transmission phase,
the BS receives the following signal in the space
and time domain:
\be
\bY = \sum_{k=1}^K \bh_k
\sqrt{P_k} \bc_{l(k)}^\rT + \bN \in \uC^{M \times N},
\ee
where $K$ represents the number of active devices,
$l(k)$ denotes the index of the preamble chosen
by active device $k$, 
$P_k$ is the transmit power of active device $k$,
and 
$[\bN]_{m,n} \sim \cC \cN(0, N_0)$ is the background noise
at the $m$th antenna and the $n$th preamble symbol duration.

Each active device transmits its 
data packet of length $D$ during the data transmission phase
(in the second sub-slot).
The corresponding received signal at the BS becomes
\be
\bZ =  \sum_{k=1}^K \bh_k
\sqrt{P_k} \bb_k^\rT + \tilde \bN \in \uC^{M \times D},
\ee
where $\bb_k$ represents the data packet from active device $k$
and 
$[\tilde \bN]_{m,d} \sim \cC \cN(0, N_0)$ is the background noise
at the $m$th antenna and the $n$th data symbol duration.
Throughout the paper, we assume that
$\uE[\bb_k] = \b0$ and $\uE[\bb_k \bb_k^\rH] = \bI$
(i.e., the symbol energy is normalized).

The BS uses the correlator to estimate the channel vector as follows:
\begin{align}
\bg_l 
& = \bY \bc_l 
= \sum_{k=1}^K \bh_k \sqrt{P_k} \delta_{l(k),l} + \bN \bc_l \cr
& = \sum_{k \in \cK_l} \bh_k \sqrt{P_k} + \bN \bc_l, 
	\label{EQ:bg_l}
\end{align}
where $\delta_{l,l^\prime}$
is the Kronecker delta (i.e.,
$\delta_{l,l^\prime} = 1$ if $l = l^\prime$, and 0 otherwise)
and $\cK_l$ represents the index set of the active
devices that choose preamble $l$.
If active device $k$ is the only 
device that chooses preamble $l$ (i.e., $l(k) = l$ or
$\cK_l  = \{k\}$),
thanks to the orthogonality of preambles,
it can be shown that
\be
\bg_l = \bh_k \sqrt{P_k} + \bn_l,
	\label{EQ:gn}
\ee
where $\bn_l = \bN \bc_l \sim \cC \cN \left(\b0, 
N_0 \bI \right)$.
To decode the data packet from 
active device $k$, conjugate
beamforming is applied and the output
of the beamformer becomes
\begin{align}
\bx_l & = \bg_l^\rH \bZ \cr
& = \bg_l^\rH \bh_k \sqrt{P_k} \bb_k^\rT +
\sum_{k^\prime \ne k} \bg_l^\rH \bh_{k^\prime} 
\sqrt{P_{k^\prime}}\bb_{k^\prime}^\rT + 
\bg_l^\rH \tilde \bN.
	\label{EQ:bd_l}
\end{align}
If $|\bg_l^\rH \bh_k|^2$ is sufficiently larger
than $|\bg_l^\rH \bh_{k^\prime}|^2$, 
a high SINR 
can be achieved for successful decoding.

\section{SINR with Estimated Channel Vectors}	\label{S:SINR}

In this section, we consider a simple expression for the SINR 
in the grant-free random access system
in Section~\ref{S:SM},
which allows tractable analysis in Section~\ref{S:Anal}.

For tractable analysis,
we consider the following assumption \cite{Bjornson16}.
\begin{itemize}
\item[{\bf A)}]
Throughout the paper,
$P_k$ is decided to be inversely proportional to
$\ell_k$ via power control so that
\be
\bh_k \sqrt{P_k} = \bv_k \sqrt{P_{\rm rx}},
        \label{EQ:A}
\ee
where $P_{\rm rx}$ represents the (average) receive signal power
and $\bv_k \sim \cC \cN(\b0, \bI)$ is independent for all
$k$ (i.e., Rayleigh fading is assumed for small-scale fading).
\end{itemize}

As in \cite{Bjornson17} \cite{Sanguinetti18},
when no power control is employed,
each device may have a different 
receive signal power at the BS that can be exploited for
collision resolution. However, as mentioned above,
we do not consider this case 
in this paper, and 
PC is declared if there are multiple active devices
that choose the same preamble, which might lead to 
a worse performance than the case where different 
receive signal powers (due to no
power control) are exploited for collision resolution.

For convenience, let
$\gamma  = \frac{P_{\rm rx}}{N_0}$ be the
signal-to-noise ratio (SNR).
For given $\{\bv_k\}$,
to find the SINR, we consider active device $k$ with 
$\cK_l = \{k\}$ (i.e., without PC).
From \eqref{EQ:bd_l}, 
it can be shown that
\begin{align}
\bx_l 
& = 
\left(P_{\rm rx} ||\bv_k||^2 + \sqrt{P_{\rm rx}} \bn_l^\rH \bv_k
\right) \bb_k^\rT \cr
& \quad + \sum_{k^\prime \ne k} (P_{\rm rx} \bv_k^\rH \bv_{k^\prime}
+ \sqrt{P_{\rm rx}}\bn_l^\rH \bv_{k^\prime}) \bb_{k^\prime}^\rT \cr
& \quad +
(\sqrt{P_{\rm rx}} \bv_k + \bn_l)^\rH \tilde \bN.
	\label{EQ:bx_l}
\end{align}
For given $\{\bv_k\}$, the conditional SINR 
of active device $k$ without PC becomes
\be
\sSINR_k = \frac{ \gamma^2 ||\bv_k||^2 + \gamma }{ I_k}, 
	\label{EQ:SINR_k}
\ee
where
\begin{align}
I_k =
\sum_{k^\prime \ne k} \left( \gamma^2
\frac{|\bv_k^\rH \bv_{k^\prime}|^2}{||\bv_k||^2} 
+ \gamma \frac{||\bv_{k^\prime}||^2}{||\bv_{k}||^2} \right) 
+ \gamma + \frac{M}{||\bv_k||^2}.
\end{align}
In \eqref{EQ:SINR_k}, 
the signal power and interference-plus-noise power terms
are obtained by taking the expectation with respect to
the noise terms, and
the term $(\sqrt{P_{\rm rx}}\bn_k^\rH
\bv_k) \bb_k^\rT$ in \eqref{EQ:bx_l}
is part of the desired signal
(i.e., $\bg_l^\rH \bh_k \sqrt{P_k} \bb_k^\rT$ in \eqref{EQ:bd_l}).
As a result, its power term, i.e., $\gamma$,
becomes the 2nd term in the numerator in \eqref{EQ:SINR_k}.

Based on the strong law of large numbers,
since $\frac{||\bv_k^\rH ||^2}{M} \to 1$ (w.p. 1) as $M \to \infty$,
it can be shown that
\begin{align}
I_k \to
\sum_{k^\prime \ne k} \gamma^2
|X_{k,k^\prime}|^2
+ (K-1) \gamma + \gamma +1,
        \label{EQ:aI_k}
\end{align}
where
$X_{k,k^\prime}
= \frac{\bv_k^\rH \bv_{k^\prime}}{||\bv_k||}$.
For given $\bv_k$, $X_{k,k^\prime}$ is seen as
a weighted sum of independent CSCG random variables.
Thus, for given $\bv_k$,
$X_{k,k^\prime}  \sim \cC \cN
\left(0, \sum_{m=1}^M \frac{|[\bv_k]_m|^2}{||\bv_k||^2} \right)$.
Since $\sum_{m=1}^M \frac{|[\bv_k]_m|^2}{||\bv_k||^2} = 1$,
we can see that
$X_{k,k^\prime} \sim \cC \cN(0, 1)$, which is independent of $\bv_k$.
Furthermore, for given $\bv_k$,
$X_{k,k^\prime}$, $k^\prime \ne k$, are independent.
Then, it can be shown that
\be
\lim_{M \to \infty} \uE[I_k] = \gamma^2 (K-1) + \gamma  K + 1,
\ee
while 
\be
\lim_{M \to \infty} \frac{\gamma^2 ||\bv_k||^2 + \gamma}{M}
= \gamma^2.
\ee
Thus, it can be shown that
\be
\lim_{M \to \infty} \frac{\uE[\sSINR_k]}{M}
= \frac{1}{ (K -1) + \frac{K + \frac{1}{\gamma}}{\gamma}}.
	\label{EQ:lim_sinr}
\ee
With a sufficiently large $M$ (but finite),
for a given $K$,
the conditional mean of SINR (with estimated
channel vectors) can be approximated as follows:
\begin{align}
\sSINR_k (K) 
= \frac{M}{K \left(1 + \frac{1}{\gamma} \right) + \frac{1}{\gamma^2}- 1} 
= \frac{M}{K b_1 + b_0},
	\label{EQ:asinr}
\end{align}
which shows that the SINR decreases with $K$.
Here,
$b_1 = 1 + \frac{1}{\gamma}$
and $b_0 = \frac{1}{\gamma^2} - 1$, which are independent of
$M$ and $K$.
Note that while the SINR expression in \eqref{EQ:asinr}
or similar ones can be found in the literature (e.g., \cite{Ngo13}
\cite{Ding19_IoT}),
it has been (re-)derived to be used in Section~\ref{S:Anal}
for the throughput and stability of the proposed approach.

\begin{figure}[thb]
\begin{center}
\includegraphics[width=\figwidth]{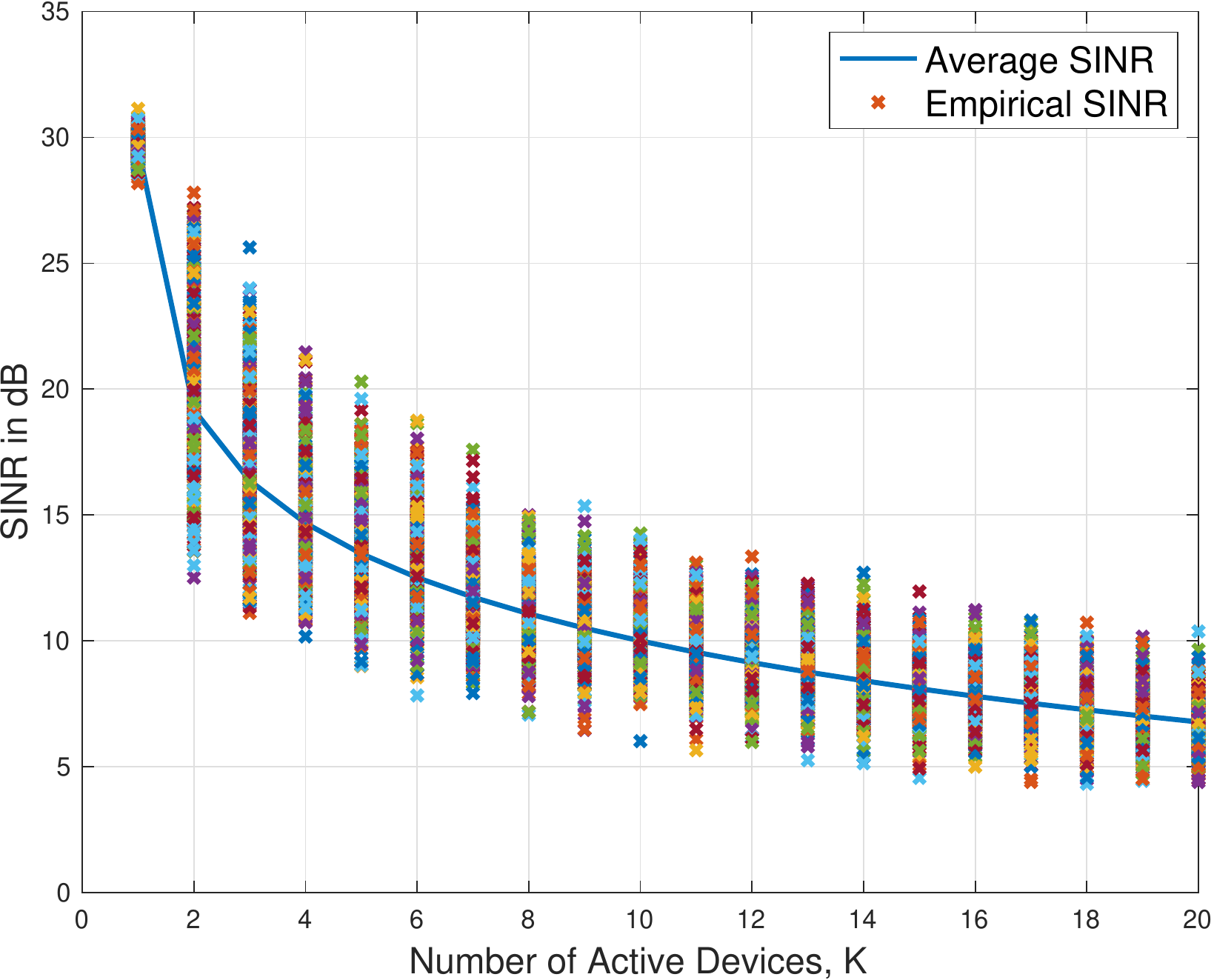}
\end{center}
\caption{Average SINR with empirical 
results (with the estimated channel
vectors) as functions of $K$ when $M = 100$ and $\gamma = 10$ dB.}
        \label{Fig:e_sinr}
\end{figure}

In Fig.~\ref{Fig:e_sinr},
empirical results of SINR
are shown with the average SINR in \eqref{EQ:asinr} 
for different values of $K$ when $M = 100$ and $\gamma = 10$ dB.
It is noteworthy that the SINR in \eqref{EQ:asinr} 
is dependent on $K$ that can be
seen as a random variable in random access.

\section{Feedback Signals and Immediate Re-Transmissions}	\label{S:IRT}

In grant-free random access with massive MIMO,
an active device can successfully transmit a packet
if \emph{i)} there is no PC \emph{and}
\emph{ii)} the SINR is sufficiently high, i.e.,
$\sSINR_k \ge \Gamma$, where $\Gamma$ is a threshold SINR
for successful decoding. 
As a result, although PC does not happen,
an active device needs to drop the packet or re-transmit 
(after a random delay) when the SINR is lower than $\Gamma$
(as decoding is unsuccessful),
which can degrade throughput.
In order to improve throughput,
re-transmission diversity (RTD) \cite{Caire01} can be employed
with two different types of feedback signals.
In this section, we discuss 
feedback signals and propose an
approach for grant-free random access with massive MIMO
based on RTD.

\subsection{Feedback after Preamble Transmission}

If there are $K$ active devices that randomly choose
the preambles in $\cC$,
the probability of PC,
which is the conditional probability that an active device
experiences PC provided that there are $K$ active devices,
is given by
\be
\uP_{\rm pc} (K) = 1 - \left(1 - \frac{1}{L} \right)^{K-1}
= 1 - \alpha_L^{K-1},
\ee
where $\alpha_L =  1 - \frac{1}{L}$. 
In general,
the channel vectors of the active devices with PC 
cannot be correctly estimated.
To see this, suppose that the $l$th preamble
is chosen by active devices 1 and 2. Then,
from \eqref{EQ:bg_l},
the output of the correlator becomes
$$
\bg_l = \bv_1 + \bv_2 + \bn_l.
$$
Thus, we can assume that the active devices
with PC have low SINRs,
while the signals from them 
become interfering signals that degrade
the SINR of 
the active devices without PC as shown in \eqref{EQ:asinr}.
For convenience, 
the active devices without PC are
referred to as PC-free (active) devices.

In order to improve the performance,
the BS can detect PCs and 
send a feedback signal after the preamble transmission phase
to inform the collided preambles. That is, for each preamble,
acknowledgment (ACK) or negative acknowledgment (NACK) 
is sent. The resulting feedback
signal is referred to as feedback signal of Type I.
With feedback signal of Type I,
the active devices with PC 
do not transmit their packets (their packets
are backlogged for re-transmissions according to
re-transmission policies or dropped).
Thus, for given $K$, 
the average number of PC-free devices
is given by
\be
\bar K (K) =  K \alpha_L^{K-1}.
\ee
In \eqref{EQ:asinr}, $K$ can be 
replaced with $\bar K (K)$, which will be simply
denoted by $\bar K$, for the SINR
if the active devices with PC 
do not transmit their data packets,
which leads to a higher SINR (because $\bar K \le K$) \cite{Ding21a}.

\subsection{Feedback after Data Packet Transmission and RTD}

For a PC-free device, decoding of data packet can fail
if its SINR is not sufficiently high.
For successful decoding, as mentioned earlier, 
the SINR has to be higher than or equal to a threshold, $\Gamma$.
When the SINR is not sufficiently high or decoding is not
successful, the data packet of a PC-free 
device can be dropped or backlogged for 
re-transmission after a random delay.
For convenience, the resulting random access system
is referred to as the conventional system.

\begin{figure}[thb]
\begin{center}
\includegraphics[width=\figwidth]{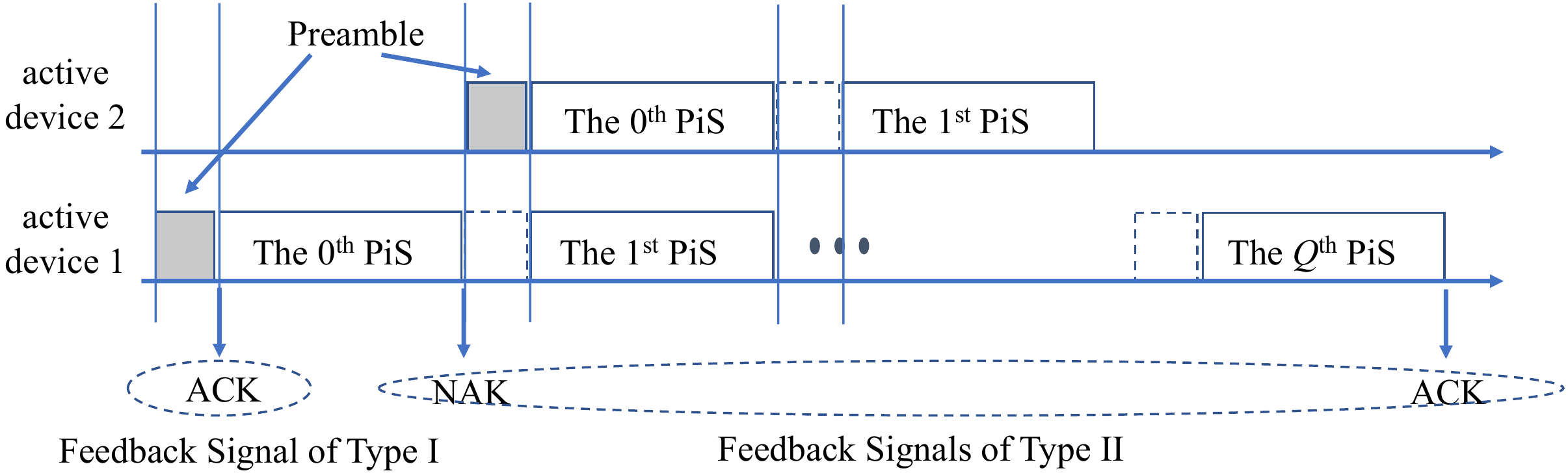}
\end{center}
\caption{An illustration 
of packet transmissions in grant-free random access with IRT
where two active devices transmit their packets with preambles.}
        \label{Fig:1}
\end{figure}

For a PC-free device, 
although its data packet cannot be decoded,
the BS knows its channel vector.
Thus, we can take advantage of known channel vector and exploit RTD 
by allowing a PC-free device to
continuously re-transmit the same packet
until decoding becomes successful to improve throughput 
as shown in Fig.~\ref{Fig:1},
where active device 1 performs $(Q+1)$-time transmissions
of a data packet.
Clearly, since its channel is estimated
by the preamble transmission in the first slot,
the PC-free device does not need to
send any preambles in the subsequent slots (this helps
reduce the PC for next active devices, e.g., active device 2
in Fig.~\ref{Fig:1}).
For convenience, the packet of a PC-free 
device that is to be re-transmitted in subsequent slots
is referred to as a packet in service (PiS).
Furthermore, the packet in the $q$th subsequent slot 
for re-transmission is referred to as the $q$th PiS.

There are few remarks as follows.

\begin{itemize}
\item  To exploit the known channels of PC-free devices,
all the re-transmissions have to be done within a
coherence time. Thus, in practice, the maximum\footnote{In this paper,
we assume that the mobility of devices in MTC is limited
and the coherence time is sufficiently long
for a number of re-transmissions.}
number of re-transmissions 
has to be limited, although it is not considered in this paper.
Furthermore, the BS can fail to decode after receiving the
maximum number of re-transmissions (e.g., due to a low 
accumulated SINR).
In this case, the BS can send 
\emph{NACK of Type I} to the device so that it can try
to send the data packet later 
(according to a re-transmission policy) or drops.

\item Due to the re-transmissions of data packet,
the SINR can increase and after a certain number of
re-transmissions,
the BS can succeed to decode the data packets.
For re-transmissions, the BS needs
to send ACK or NACK
feedback signal (for success or 
failure of decoding), which is referred to as
Feedback Signal of Type II in this paper.
In addition, the resulting random access system
is referred to as the immediate re-transmission (IRT) based system
or IRT system.

\item Since a PC-free device in the IRT system 
can transmit more than one data packet (i.e., multiple PiSs),
the SINR can be low
and a lower SINR results in more re-transmissions.
As a result, the IRT system can be unstable
(i.e., the number of re-transmissions becomes unbounded).
In the next section, we address the stability
issue of IRT through throughput analysis.
\end{itemize}

\section{Throughput Analysis}	\label{S:Anal}

In this section, we study the throughput
of the IRT system. For tractable analysis,
we assume that the arrivals at each slot
are independent and follow a Poisson distribution
with mean arrival rate $\lambda$ (in the number of 
new active devices per slot),
i.e.,
\be
K \sim {\rm Pois} (\lambda).
	\label{EQ:KPois}
\ee
Throughout this section, we assume that
an active device drops its data packet when it receives NACK of Type I.

\subsection{Conventional System}

In the conventional system, for given $K$ in a slot,
the average number of dropped packets
due to PC as well as low SINR
is $\uE[U(K)\,|\,K]$, where
\begin{align}
U (K) = K \left(1 - \alpha_L^{K-1}
\right) + \bar K \indicator (\sSINR_k < \Gamma).
	\label{EQ:U_K}
\end{align}
In \eqref{EQ:U_K}, in order to obtain
the mean of the second term on the right-hand side (RHS),
the distribution of the SINR is required.
Although it might be possible to find the distribution of the SINR,
for tractable analysis, we use the conditional
mean of SINR in \eqref{EQ:asinr}, which is a function of $\bar K$,
in this subsection.

Assume that $\bar K$ is a Poisson random variable with 
the following mean:
\begin{align}
\bar \lambda = \uE[\bar K] 
= \sum_{k=0}^\infty k \alpha_L^{k-1} p_\lambda (k)
= \lambda e^{- \frac{\lambda}{L}},
\end{align}
where $p_\lambda (k) = \frac{e^{-\lambda} \lambda^k}{k!}$.
Then,
it can be shown that
\begin{eqnarray}
\uE[ \bar K \indicator (\sSINR_k(\bar K) < \Gamma)] 
& = &\sum_{k=0}^\infty k \indicator 
\left( \frac{M}{k b_1 + b_0} < \Gamma \right) p_{\bar \lambda}(k) \cr
& = &\sum_{k > K_\Gamma} k p_{\bar \lambda}(k) \cr
& = &\bar \lambda -  
\sum_{k=0}^{\lfloor K_\Gamma
\rfloor} k p_{\bar \lambda}(k) \cr
& = & \bar \lambda - \bar 
F_{\bar \lambda} (\lfloor K_\Gamma \rfloor -1),
	\label{EQ:bKi}
\end{eqnarray}
where 
$F_{\bar \lambda}(n) =
\sum_{k=0}^{n} \frac{ e^{-\bar \lambda} \bar \lambda^k}{k!}$ is the 
cumulative distribution function (cdf) with mean $\bar \lambda$ and
\be
K_\Gamma = 
\frac{1}{b_1} \left( \frac{M}{\Gamma} - b_0\right).
\ee

Clearly, $K_\Gamma$ is a constant that is independent of $K$.
From \eqref{EQ:U_K} and \eqref{EQ:bKi},
it can be shown that
\begin{align}
\lambda_{\rm d} & \deft \uE[U (K)] = \uE[\uE[U (K)\,|\, K]] \cr
& =  \lambda - \bar \lambda + \bar \lambda 
-\bar \lambda e^{-\bar \lambda} \sum_{k=0}^{\lfloor K_\Gamma \rfloor - 1} 
\frac{ \bar \lambda^k}{k!} \cr
& =  \lambda - 
\bar \lambda 
e^{-\bar \lambda} \sum_{k=0}^{\lfloor K_\Gamma \rfloor - 1} 
\frac{ \bar \lambda^k}{k!}.
	\label{EQ:lb1}
\end{align}
Let
\be
\lambda_{\rm n} = \lambda - \lambda_{\rm d} = 
\bar \lambda F_{\bar \lambda}(
\lfloor K_\Gamma  \rfloor - 1),
	\label{EQ:ln1}
\ee
where $\lambda_{\rm n}$ represents the average
number of successfully transmitted packets per slot
or throughput.
Thus, the maximum throughput of the conventional system
becomes
\begin{align}
\eta_{\rm con} & = 
\max_{\lambda} \bar \lambda F_{\bar \lambda}(
\lfloor K_\Gamma  \rfloor - 1) \cr
& = \max_{0 \le \bar \lambda \le L e^{-1}} \bar \lambda F_{\bar \lambda}(
\lfloor K_\Gamma  \rfloor - 1),
	\label{EQ:e_con}
\end{align}
because $\bar \lambda \le \max_\lambda \lambda e^{-\frac{\lambda}{L}}
= L e^{-1}$.
In addition, let 
$$
\lambda_{\rm con} = 
\argmax_{\lambda} \bar \lambda F_{\bar \lambda}(
\lfloor K_\Gamma  \rfloor - 1).
$$
The maximum throughput of the 
conventional system can approach $L e^{-1}$ if $K_\Gamma$ is sufficiently
large. Clearly, for a fixed $\Gamma$, as $M \to \infty$,
$K_\Gamma \to \infty$, which implies
that $\eta_{\rm con} \to L e^{-1}$,
i.e., the performance of the 
conventional system with a sufficiently
large $M$, is limited by the number of preambles, $L$.
However, in practice, since $M$ is finite,
it is expected that $\eta_{\rm con}$ is
lower than $L e^{-1}$.

\subsection{Immediate Re-Transmission based System}

In the IRT system,
only PC
results in dropped packets.
Thus, for given $K$, the average number of dropped packets
is given by $U(K) = K (1 - \alpha_L^{K-1})$.
From this, we have
\begin{align}
\lambda_{\rm d} = \uE[U(K)] 
= \lambda (1 - e^{-\frac{\lambda}{L}}) = \lambda - \bar \lambda.
	\label{EQ:lb2}
\end{align}
Clearly, from \eqref{EQ:lb1}
and \eqref{EQ:lb2}, 
it can be shown that
\be
\lambda_{\rm n} = \bar \lambda = \lambda e^{-\frac{\lambda}{L}}  \le L e^{-1},
	\label{EQ:ln2}
\ee
and the maximum throughput of the IRT system can be
\be
\eta_{\rm irt} = \max_\lambda \bar \lambda = L e^{-1},
	\label{EQ:e_irt}
\ee
where the upper-bound is achieved when $\lambda = L$.
That is, in the IRT system, with a finite $M$,
it seems that a maximum throughput of $L e^{-1}$ is achievable.
However, if there are a large
number of PC-free devices, 
the SINR becomes low and the number of PiSs increases,
which may make the IRT system unstable.
Thus, for a stable IRT system, it is expected that
any PC-free devices receive
ACK of Type II after a \emph{finite} number of re-transmissions.
Thus, we now focus on 
the derivation of 
key conditions for a finite number of PiSs 
and the throughput of stable IRT.

Let $A(t)$ denote the number of new data packets
arrived at slot $t$. That is,
$A(t)$ is the number of new PC-free devices in slot $t$.
Then, it can be shown that
$\uE[A(t)] = \uE[K \alpha_L^{K-1}] = \bar \lambda$,
i.e., 
\be
A(t) \sim {\rm Pois} (\bar \lambda).
	\label{EQ:At}
\ee
We assume that $A(t)$ is an independent Poisson random
variable with mean  $\lambda - \bar \lambda$.
In addition, denote by $\bar K(t)$ the total number of the
data packets in slot $t$.
In the IRT system,
denote by $W_{q,t^\prime}$ 
the number of the $q$th PiS that is first transmitted at time slot
$t^\prime$. 
Thus, the number of the $q$th PiS at time slot $t$ is $W_{q, t-q}$,
and it can be shown that
\be
\bar K(t) = A(t) + \sum_{q=1}^\infty W_{q,t-q}.
	\label{EQ:KA}
\ee
Note that in the conventional system, $\bar K (t) = A(t)$ as
no immediate re-transmissions are considered.

Since all the PiSs are copies of the 0th PiS,
the BS can exploit the 
RTD gain by combining all the PiSs 
and the resulting SINR is a sum of the SINRs of individual PiSs
\cite{Caire01}. Thus,
in slot $t$, the SINR 
of a data packet from a PC-free device
with $(q+1)$-time transmissions is given by
\begin{align}
\sSINR_{\rm rtd} 
& = \frac{M}{\bar K(t) b_1 + b_0}
+ \ldots + \frac{M}{\bar K(t-q) b_1 + b_0} \cr
& = \frac{M}{\bar K(t) b_1 + b_0}
\left(1 +
\sum_{i=1}^q
\frac{\bar K(t) b_1 + b_0}{\bar K(t-i) b_1 + b_0}
\right) \cr
& \approx \sSINR(\bar K (t),q),
	\label{EQ:rtd_sinr}
\end{align}
where
\be
\sSINR (K, q) = \frac{(q+1) M}{K b_1 + b_0}.
	\label{EQ:rtd_sinr2}
\ee
Here, the approximation
in \eqref{EQ:rtd_sinr} is valid if $\bar K(t)$
is not significantly varying over $t$,
i.e., $\bar K(t) \approx \cdots \approx \bar K (t- q)$.
With RTD,
it is clear that the SINR of PiS increases with $q$
provided that $K$ is fixed or bounded.
Thus, for a sufficiently large $q$, the packet can be decoded
and $W_{q, t-q}$ becomes 0.
In other words, 
the total number of packets in slot $t$, $\bar K(t)$ in
\eqref{EQ:KA},
can be bounded.

\begin{mylemma}
With RTD, 
suppose that the SINR of a packet with $(q+1)$th transmissions
is given by \eqref{EQ:rtd_sinr2}.
Then, the number of PiSs in slot $t$
becomes
\begin{align}
\bar K(t) = A(t) + \sum_{q=1}^{Q_\Gamma (\bar K(t-1))} A(t-q),
	\label{EQ:L1}
\end{align}
where
\begin{align}
Q_\Gamma (K) = \min_q \{ q\,:\, \sSINR (K, q) \ge \Gamma \}.
	\label{EQ:Q_G}
\end{align}
\end{mylemma}
\begin{IEEEproof}
Since 
$\sSINR (K,q) > \sSINR (K, q-1)$,
we can show that
\be
\sSINR(K, Q_\Gamma (K)) \ge \Gamma \ \mbox{and}\ 
\sSINR(K, Q_\Gamma (K)-1) < \Gamma.
	\label{EQ:SSG}
\ee
Thus, according to \eqref{EQ:SSG}, in slot $t-1$, 
the $Q_\Gamma (\bar K (t-1))$th PiS is decodable, but
the $(Q_\Gamma (\bar K (t-1))-q)$th PiS is not,
where $q = 1, \ldots, Q_\Gamma (\bar K (t-1))$.
As a result, in slot $t$, 
we can show that
$$
\sum_{q=1}^\infty W_{q,t-q} = 
\sum_{q=1}^{(Q_\Gamma (\bar K(t-1))-1) +1} W_{q, t-q}
$$
or
\begin{align}
\bar K(t) = A(t) + \sum_{q=1}^{Q_\Gamma (\bar K(t-1))} W_{q,t-q}.
	\label{EQ:KAW}
\end{align}

Since the number of the $0$th PiSs
in slot $t$ is $A(t)$, i.e.,  $W_{0, t} = A(t)$,
$W_{q, t-q}$ is either $A(t-q)$ or 0.
That is, $W_{q,t-q}$ can be
replaced with $A(t-q)$ unless the $q$th PiSs are decodable.
Thus, letting $W_{q,t-q} = A(t-q)$ in \eqref{EQ:KAW},
we can obtain \eqref{EQ:L1}, which completes the proof.
\end{IEEEproof}

From \eqref{EQ:L1},
it can be shown that
if $\bar K (t)$ increases with $t$,
it results in the decrease of SINR and the increase
of $Q_\Gamma (\bar K (t))$ based on the definition
of $Q_\Gamma (K)$ in \eqref{EQ:Q_G}. 
This implies that $\bar K(t)$ can be unbounded and
the IRT system becomes unstable.
To avoid it, it is necessary to find stability conditions.
To this end, we consider Foster-Lyapunov criteria
\cite{Kelly_Yudovina} \cite{HajekBook}.

In \eqref{EQ:L1},
since $\bar K(t)$ depends on $\bar K(t-1)$,
it can be seen as a Markov chain. Since $A(t) \in \{0, 1, \ldots\}$,
the state space of $\bar K(t)$ becomes
$$
\bar K (t) \in \uZ^* \deft \{0, 1, \ldots\},
$$
where $\uZ^*$ is the set of non-negative integers.

\begin{mylemma}
If 
\be
M > \bar \lambda \Gamma b_1 
= \lambda e^{-\frac{\lambda}{L}} 
\Gamma
\left(1+ \frac{1}{\gamma} \right),
	\label{EQ:L2}
\ee
$\bar K (t)$ in \eqref{EQ:L1}
is a positive recurrent Markov chain.
\end{mylemma}
\begin{IEEEproof}
Since $\bar K (t) \ge 0$, we
assume that $\bar K(t)$ itself is a Lyapunov function.
From \eqref{EQ:Q_G},
\be
Q_\Gamma (K) = \lceil \frac{\Gamma (K b_1 + b_0)}{M} - 1 \rceil.
	\label{EQ:Qg1}
\ee
From \eqref{EQ:At}, \eqref{EQ:L1}, and \eqref{EQ:Qg1},
we have
\begin{align}
\uE[\bar K(t) \,|\, \bar K(t-1) = k]
& = \bar \lambda \left(1 + Q_\Gamma (k)  \right) \cr
& = \bar \lambda \left(\frac{\Gamma (k b_1 + b_0)}{M} + \delta
\right) \cr
& = \frac{\bar \lambda \Gamma b_1}{M} k +
\frac{\bar \lambda  \Gamma b_0}{M} + \delta,
	\label{EQ:Ckk}
\end{align}
where $\delta \in [0, 1)$,
since $\lceil x \rceil$ can be expressed as $x + \delta$.

Suppose that \eqref{EQ:L2} holds.
Then, since $\frac{\bar \lambda \Gamma b_1}{M}  > 1$,
there exists a finite $k^*$ as follows:
\be
k^* =  \min_k \left\{k \,:\,
 \frac{\bar \lambda \Gamma b_1}{M} k +
\frac{\bar \lambda  \Gamma b_0}{M}  > k \right\}.
\ee
Let
\be
\cK = \{0, 1, \ldots, k^*\},
\ee
which is a finite set. Then, 
for $\epsilon > 0$, it can be shown that
\begin{align}
d(k) & = \uE[\bar K(t) \,|\, \bar K(t-1) = k] - k < - \epsilon,
\ k \in \cK^c \cr
d(k) & < 
\frac{\bar \lambda  \Gamma b_0}{M} + 1 - \epsilon, \ k \in \cK,
	\label{EQ:dd}
\end{align}
where $\cK^c = \uZ^* \setminus \cK$.
According to \cite[Proposition D.1]{Kelly_Yudovina},
\eqref{EQ:dd} implies that
$\bar K(t)$ is a positive recurrent Markov chain.
\end{IEEEproof}

Consequently, with stable IRT, 
from \eqref{EQ:L2}, the maximum throughput 
in \eqref{EQ:e_irt} is to be replaced by
\be
\eta_{\rm irt} = \min \left\{ L e^{-1},
\frac{M }{\Gamma \left(1 + \frac{1}{\gamma} \right)} \right\}.
	\label{EQ:e_irt2}
\ee
The arrival rate corresponding to $\eta_{\rm irt}$
is denoted by $\lambda = \lambda_{\rm irt}$.
Note that the throughput of IRT is $\bar \lambda$
as mentioned earlier as long as \eqref{EQ:L2} holds.

\begin{figure}[thb]
\begin{center}
\includegraphics[width=\figwidth]{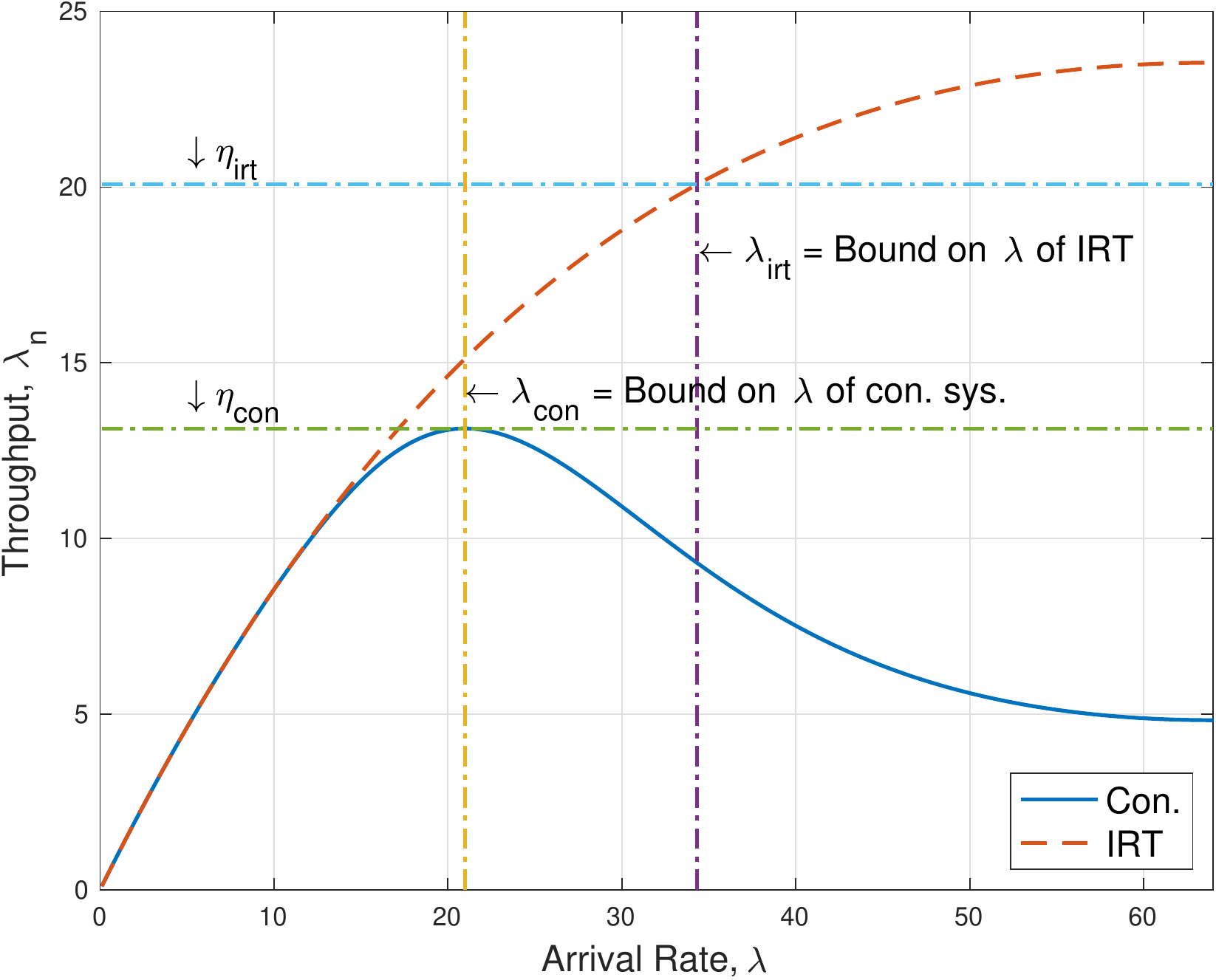}
\end{center}
\caption{Throughput, $\lambda_n$, as a function of $\lambda$
for the conventional and IRT systems
when $M = 100$, $L = 64$, and $\gamma = \Gamma = 6$ dB.}
        \label{Fig:com_lam2}
\end{figure}

In Fig.~\ref{Fig:com_lam2}, we show 
$\lambda_{\rm n}$ as a function of $\lambda$
for the conventional and IRT systems
when $M = 100$, $L = 64$, and $\gamma = \Gamma = 6$ dB.
In this case, $\lfloor K_\Gamma \rfloor$ is 20.
In addition, it can be found that
$\eta_{\rm con} = 13.13$ and $\eta_{\rm irt} = 20.07$,
i.e., the IRT system has a higher throughput
than the conventional system.
Note that in the IRT system, if $\lambda > \lambda_{\rm irt}$,
it becomes unstable (i.e., $\bar K(t) \to \infty$).

\begin{mylemma}
If \eqref{EQ:L2} holds,
\be
\eta_{\rm irt} \ge \eta_{\rm con}.
	\label{EQ:ee}
\ee
\end{mylemma}
\begin{IEEEproof}
From \eqref{EQ:e_con}, 
we have
\be
\eta_{\rm con} \le \bar  \lambda.
	\label{EQ:ecll}
\ee
From \eqref{EQ:e_irt2},  $\eta_{\rm irt}$ is either 
$\frac{M}{\Gamma b_1}$ or $L e^{-1}$.
If $\eta_{\rm irt} = L e^{-1}$, since $\bar \lambda = L e^{-1}$,
from \eqref{EQ:ecll}, we have \eqref{EQ:ee}.
On the other hand, suppose that
$\eta_{\rm irt} \to \frac{M}{\Gamma b_1}$.
If \eqref{EQ:L2} holds, we have $\frac{M}{\Gamma b_1} > \bar \lambda$,
which implies \eqref{EQ:ee}. This completes the proof.
\end{IEEEproof}

\begin{figure}[thb]
\begin{center}
\includegraphics[width=\figwidth]{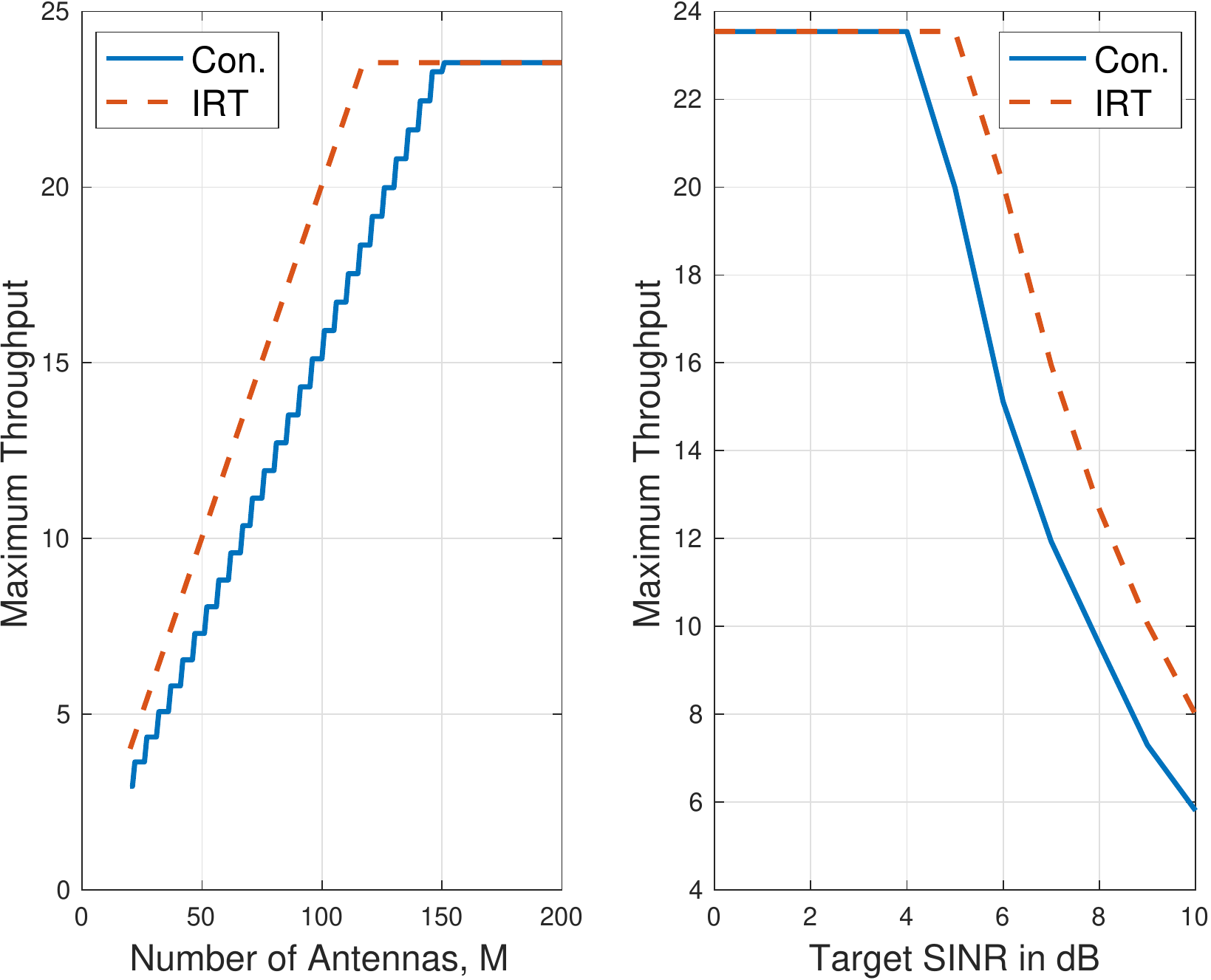} \\
\hskip 0.5cm (a) \hskip 3.5cm (b) \\
\end{center}
\caption{The maximum throughputs of conventional and IRT
systems: 
(a) throughput versus $M$ 
with $L = 64$ and $(\Gamma, \gamma) = (6, 6)$ in dB;
(b) throughput versus $\Gamma$ 
with $M = 100$, $L = 64$ and $\gamma = 6$ dB.}
        \label{Fig:eta_two}
\end{figure}

In Fig.~\ref{Fig:eta_two},
the maximum throughputs of the conventional 
and the IRT systems are shown as functions
of $M$ and $\Gamma$
when $L = 64$ and $\gamma = 6$ dB.
Unless $M$ is sufficiently large (for a fixed $\Gamma$)
or $\Gamma$ is sufficiently low (for a fixed $M$), 
it is shown that $\eta_{\rm irt}$ is higher than
$\eta_{\rm con}$. However, for a large $M/\Gamma$,
both the systems have the same maximum throughput,
$L e^{-1}$, which is limited by the number of preambles.

Note that 
since $\bar \lambda \le L e^{-1}$,
a sufficient condition for \eqref{EQ:L2} can be found as
\be
M > L e^{-1} \Gamma b_1,
	\label{EQ:SL2}
\ee
which is independent of $\lambda$. 
In addition, since $\bar \lambda \le \lambda$,
another sufficient condition is given by
\be
M > \lambda \Gamma b_1,
	\label{EQ:S2}
\ee
which is independent of $L$.

Since $\bar K(t)$ is a positive 
recurrent Markov chain when \eqref{EQ:L2} holds,
$\bar K(t)$ has a unique stationary distribution \cite{Norris98}
and $\mu_{\bar K} = \uE[\bar K(t)]$ exists.
Let $\pi (k) = \Pr(\bar K(t) = k)$ be the stationary
distribution.
From \eqref{EQ:L1},
it can be shown that
\begin{align}
\mu_{\bar K} = \sum_{k} k \pi (k) 
= \bar \lambda  \left(1 + \sum_k Q_\Gamma (k) \pi (k) \right).
	\label{EQ:mk1}
\end{align}
Unfortunately, it is difficult to find
$\mu_{\bar K}$ as $Q_\Gamma (k)$ is a nonlinear function
of $k$. However, with an approximation of
the ceiling function, i.e., $\lceil x \rceil \approx x + \frac{1}{2}$,
$Q_\Gamma (k)$ in \eqref{EQ:Qg1} can be approximated as 
$$
Q_\Gamma (k) \approx \frac{\Gamma (k b_1 + b_0)}{M} - 1 + \frac{1}{2},
$$ 
which can be used in \eqref{EQ:mk1} to find 
the following approximation of $\mu_{\bar K}$:
\be
\mu_{\bar K}
\approx \frac{
\bar \lambda \left(\frac{M}{2} +\Gamma b_0
\right)}{M - \bar \lambda \Gamma b_1}.
	\label{EQ:amu}
\ee
We can also have bounds on $\mu_{\rm K}$ as follows.

\begin{mylemma}
With $\bar K(t)$ is in \eqref{EQ:L1},
we have
\begin{align}
\bar \lambda \le \mu_{\bar K} \le 
\frac{\bar \lambda(M+\Gamma b_0)}{M - \bar \lambda \Gamma b_1}.
	\label{EQ:L3}
\end{align}
\end{mylemma}
\begin{IEEEproof}
From \eqref{EQ:Ckk}, since $Q_\Gamma (K) \ge 0$,
we can have the lower bound in \eqref{EQ:L3}
with $Q_\Gamma(K) = 0$. For the upper-bound,
it can be shown that
\begin{align}
\uE[\bar K(t)\,|\, \bar K(t-1)]
\le \frac{\bar \lambda \Gamma b_1}{M} \bar K (t-1)
+ \frac{\bar \lambda \Gamma b_0}{M} + 1.
\end{align}
Then, we have
\begin{align}
\mu_{\bar K} 
& = \uE[\uE[\bar K(t)\,|\, \bar K(t-1)]] \cr
& \le \frac{\bar \lambda \Gamma b_1}{M} \mu_{\bar K}
+ \frac{\bar \lambda \Gamma b_0}{M} + 1,
	\label{EQ:L3ll}
\end{align}
which results in the upper bound in 
\eqref{EQ:L3}.
\end{IEEEproof}

According to Little's law \cite{Kelly_Yudovina},
the average number of re-transmissions till ACK of Type II
for a PC-free device is bounded
as follows:
\begin{align}
1 \le \tau_{\rm PiS} 
= \frac{ \mu_{\bar K}}{\bar \lambda} 
	\label{EQ:xx}
\le 
\frac{M +\Gamma b_0}{M - \bar \lambda \Gamma b_1}.
\end{align}
Using the approximation in \eqref{EQ:amu},
the approximate average number of re-transmissions 
can also be obtained as follows:
\be
\tilde \tau_{\rm PiS} = 
\max\left\{1,
\frac{\frac{M}{2} +\Gamma b_0}{M - \bar \lambda \Gamma b_1} \right\}.
	\label{EQ:yy}
\ee
For a short access delay, clearly,
a small $\tau_{\rm PiS}$ is desirable.

\section{Simulation Results}	\label{S:Sim}

In Section~\ref{S:Anal},
we analyze the conventional and IRT systems
with various assumptions and approximations.
In particular, the approximations of SINR
play a key role in simplifying the analysis
(e.g., \eqref{EQ:bKi}) 
and finding the condition for stable IRT in \eqref{EQ:L3}
(e.g., \eqref{EQ:rtd_sinr}).
In this section, 
we present simulation results and show that the analysis
results agree with them.
In simulations, new arrivals are generated according to
\eqref{EQ:KPois} and the $\bv$'s
are generated as independent CSCG random vectors
according to the assumption of {\bf A)}.
The actual SINR in \eqref{EQ:SINR_k} is used
for successful decoding in both the systems and
immediate re-transmissions in the IRT system.

\begin{figure}[thb]
\begin{center}
\includegraphics[width=\figwidth]{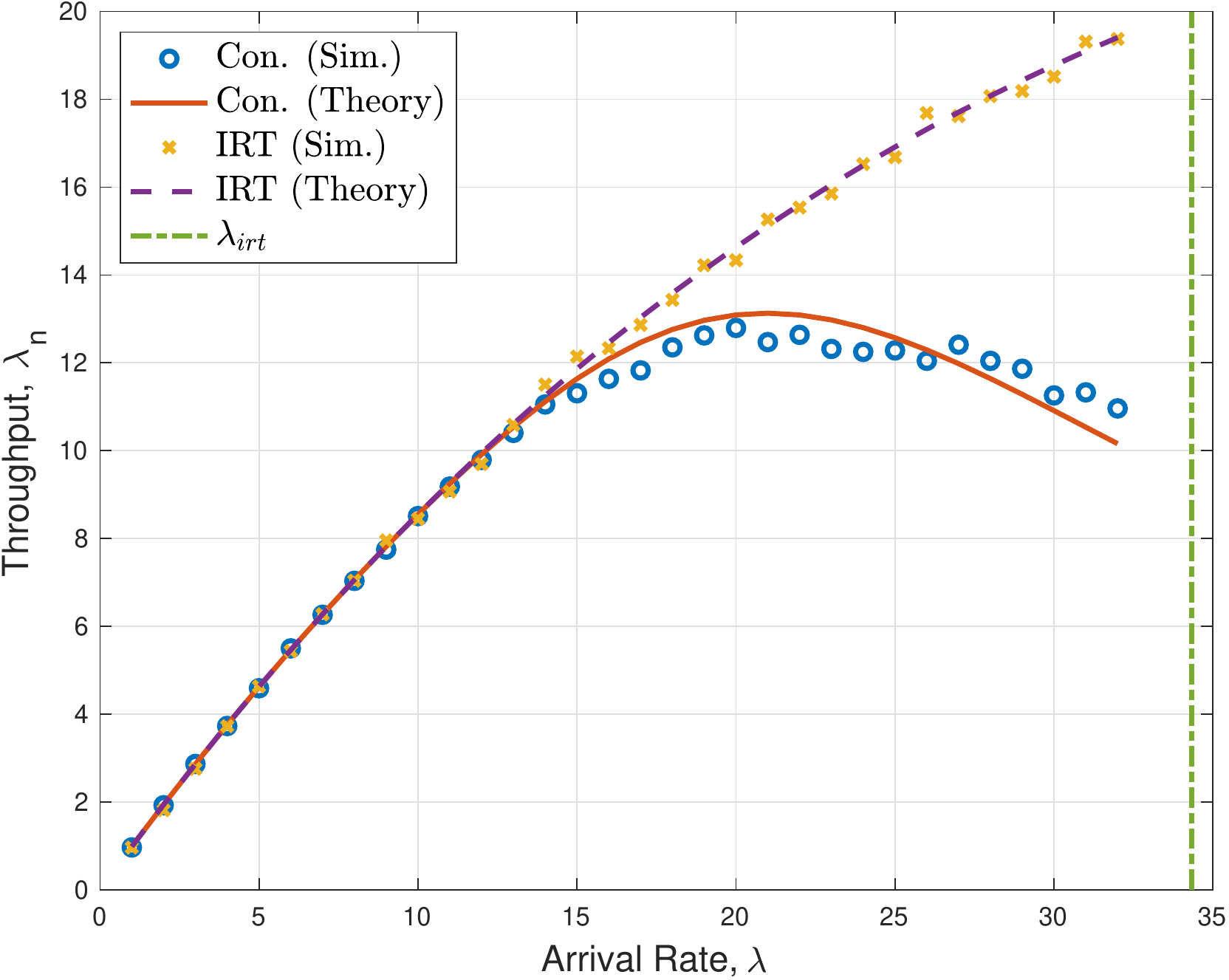} \\
(a) \\
\includegraphics[width=\figwidth]{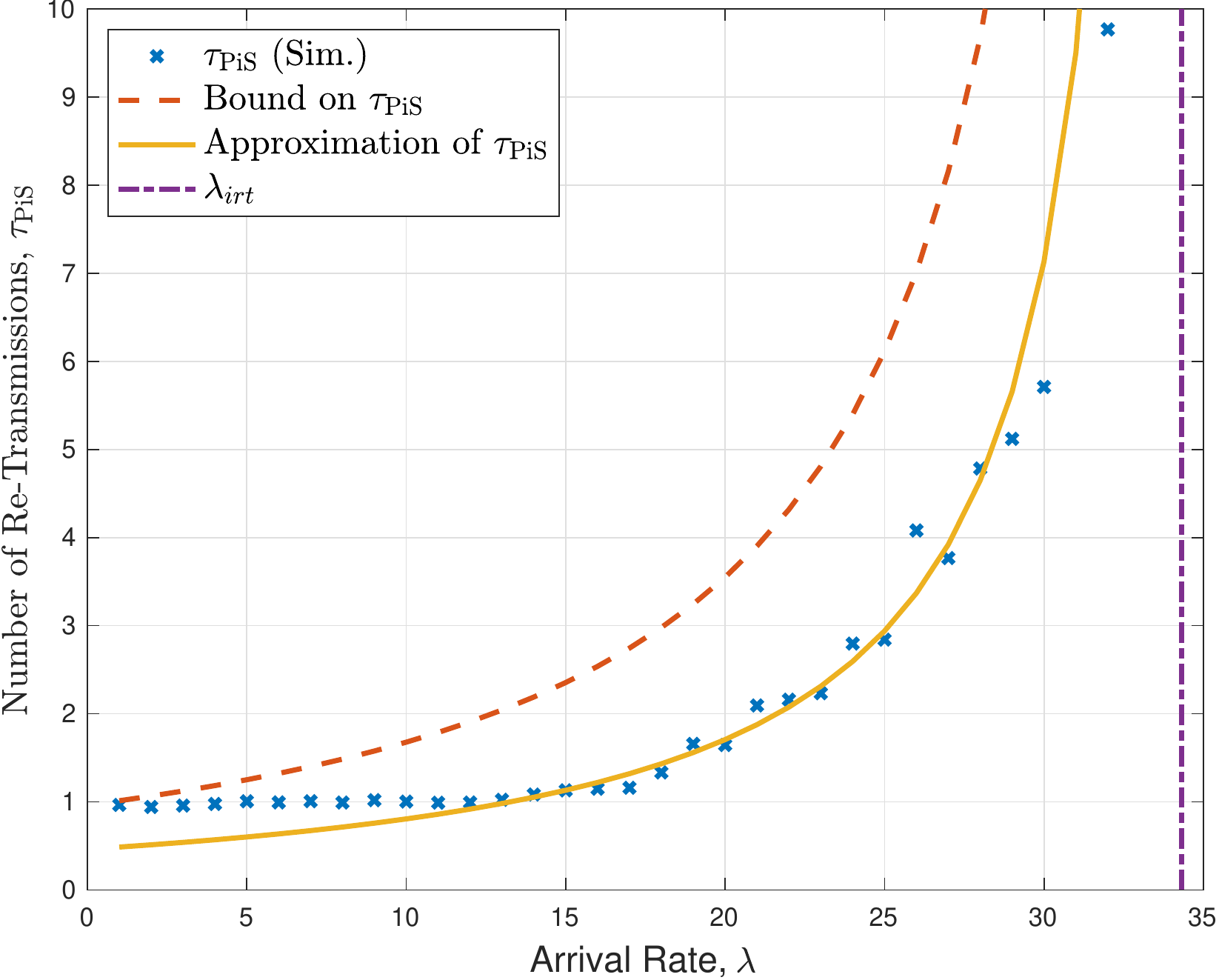} \\
(b) \\
\end{center}
\caption{Performance of the conventional
and IRT systems for different values of arrival rate, $\lambda$,
when $M = 100$, $L = 64$, and $\gamma = \Gamma = 6$ dB:
(a) throughput versus $\lambda$;
(b) $\tau_{\rm PiS}$ versus $\lambda$.}
        \label{Fig:plt_thp1}
\end{figure}

Fig.~\ref{Fig:plt_thp1} (a) shows the throughputs of 
the conventional and IRT systems
for different values of arrival rate, $\lambda$,
when $M = 100$, $L = 64$, and $\gamma = \Gamma = 6$ dB.
Thanks to RTD in the IRT system,
we can see that the throughput of the IRT system
is higher than that of the conventional system.
In Fig.~\ref{Fig:plt_thp1} (b),
it is shown that
$\tau_{\rm PiS}$ of the IRT system increases
with $\lambda$, and as $\lambda$ approaches $\lambda_{\rm irt}$,
$\tau_{\rm PiS} \to \infty$ (i.e., the IRT system becomes
unstable).
Consequently, $\lambda$ has to be lower
than $\lambda_{\rm irt}$.
Note that
the upper-bound on the and approximation of $\tau_{\rm PiS}$
shown in Fig.~\ref{Fig:plt_thp1} (b)
are obtained from \eqref{EQ:xx} and \eqref{EQ:yy}, respectively.

\begin{figure}[thb]
\begin{center}
\includegraphics[width=\figwidth]{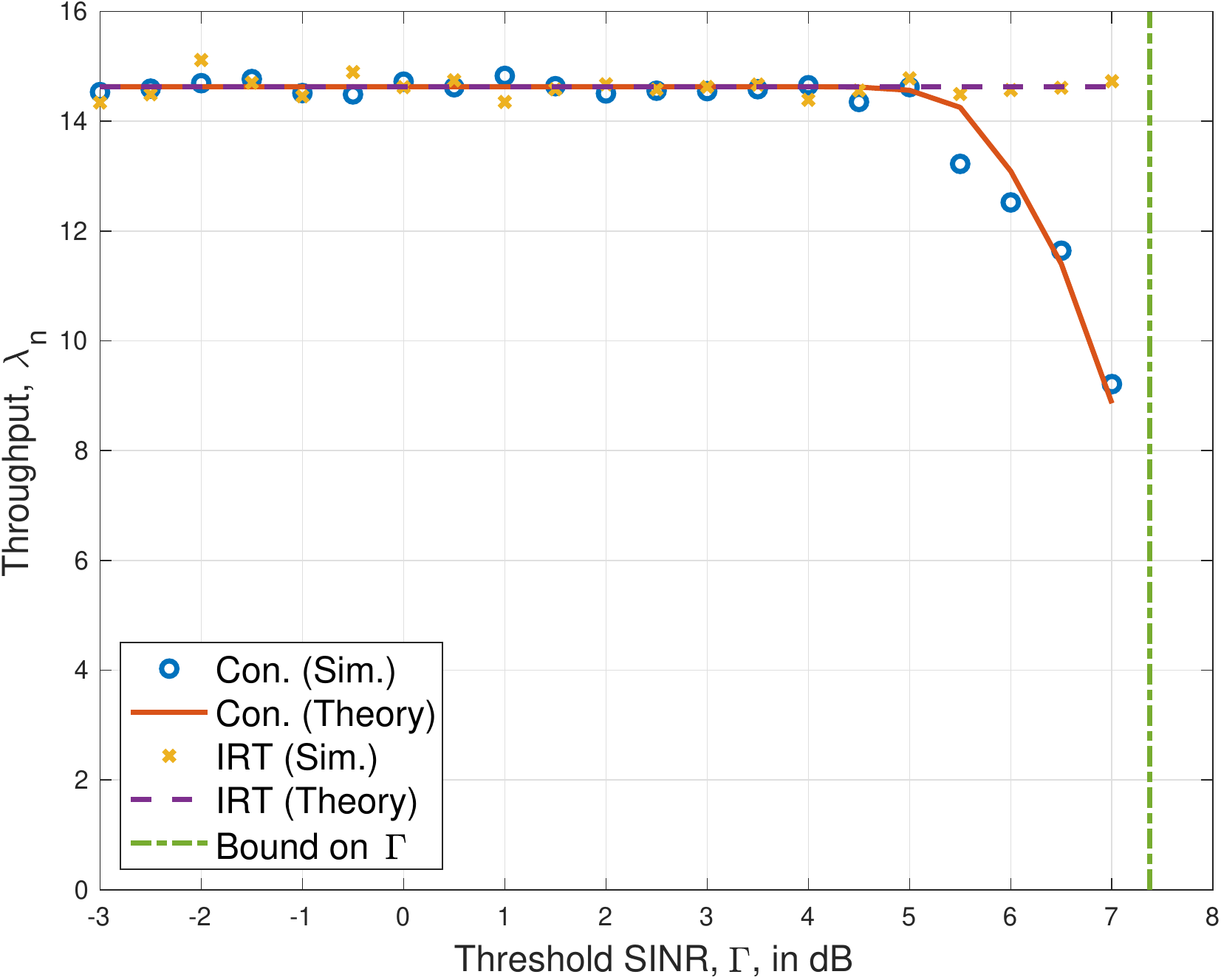} \\
(a) \\
\includegraphics[width=\figwidth]{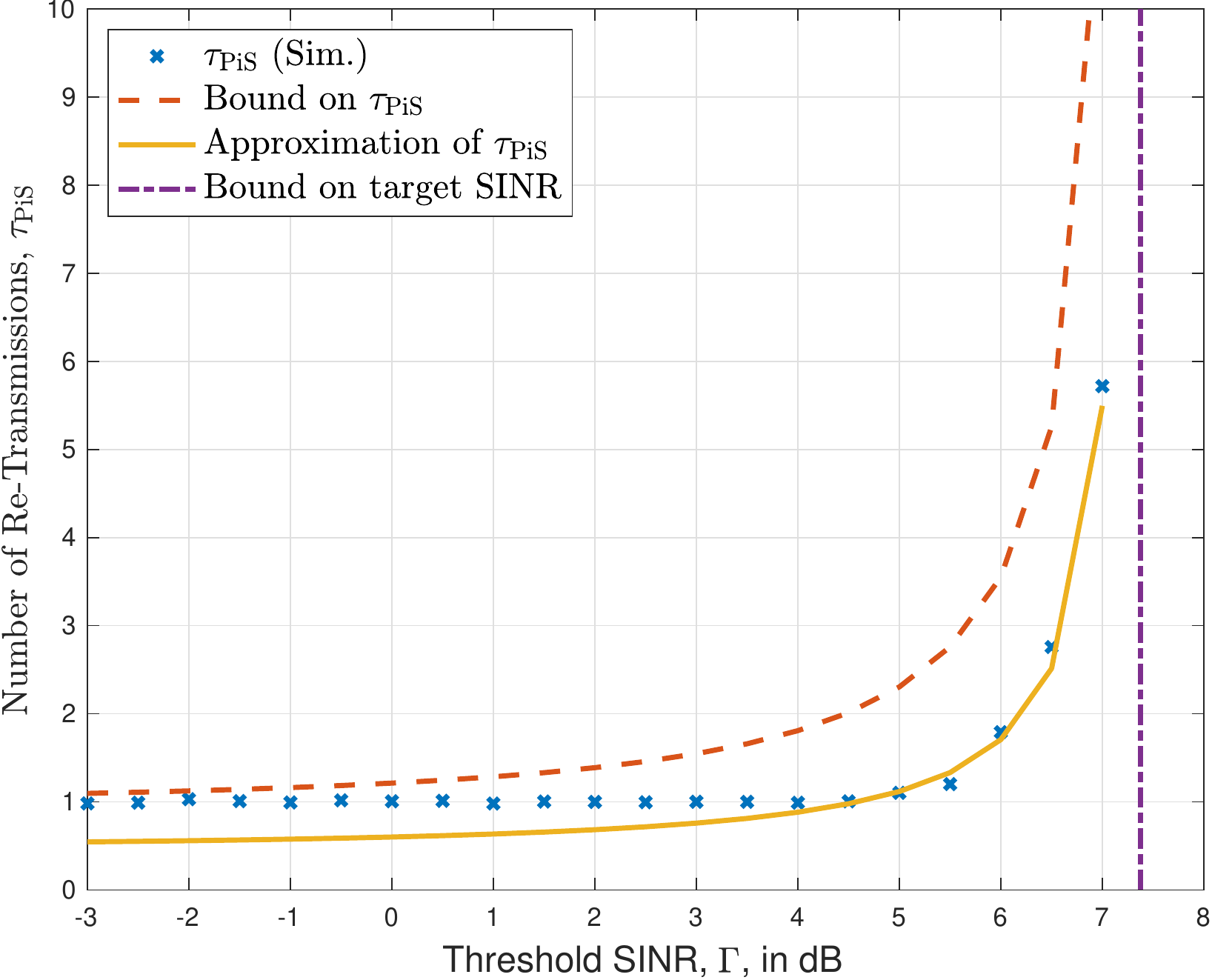} \\
(b) \\
\end{center}
\caption{Performance of the conventional
and IRT systems for different values of threshold SINR, $\Gamma$,
when $M = 100$, $L = 64$, and $\gamma = \Gamma = 6$ dB:
(a) throughput versus $\Gamma$; (b) $\tau_{\rm PiS}$ versus $\Gamma$.}
        \label{Fig:plt_thp2}
\end{figure}

In 
Fig.~\ref{Fig:plt_thp2} (a), the throughputs of 
the conventional and IRT systems
for different values of threshold SINR, $\Gamma$,
are shown 
when $M = 100$, $L = 64$, $\lambda = 20$, and $\gamma = 6$ dB.
The bound on $\Gamma$ is $\frac{M}{\bar \lambda b_1}
= 7.37$ (dB) from \eqref{EQ:L2}.
In the IRT system, with $\Gamma <
\frac{M}{\bar \lambda b_1}$, its throughput
is $\bar \lambda = \lambda e^{-\lambda/L} = 14.63$.
However, the throughput of the
conventional system becomes lower than $14.63$ when $\Gamma > 4$ dB.
Fig.~\ref{Fig:plt_thp2} (b) shows
$\tau_{\rm PiS}$ as a function of $\Gamma$.
As $\Gamma$ approaches 
$\frac{M}{\bar \lambda b_1}$,
we can see that $\tau_{\rm PiS}$ approaches $\infty$.
Thus, for stable IRT, $\Gamma$, has to be lower than 
$\frac{M}{\bar \lambda b_1}$.

\begin{figure}[thb]
\begin{center}
\includegraphics[width=\figwidth]{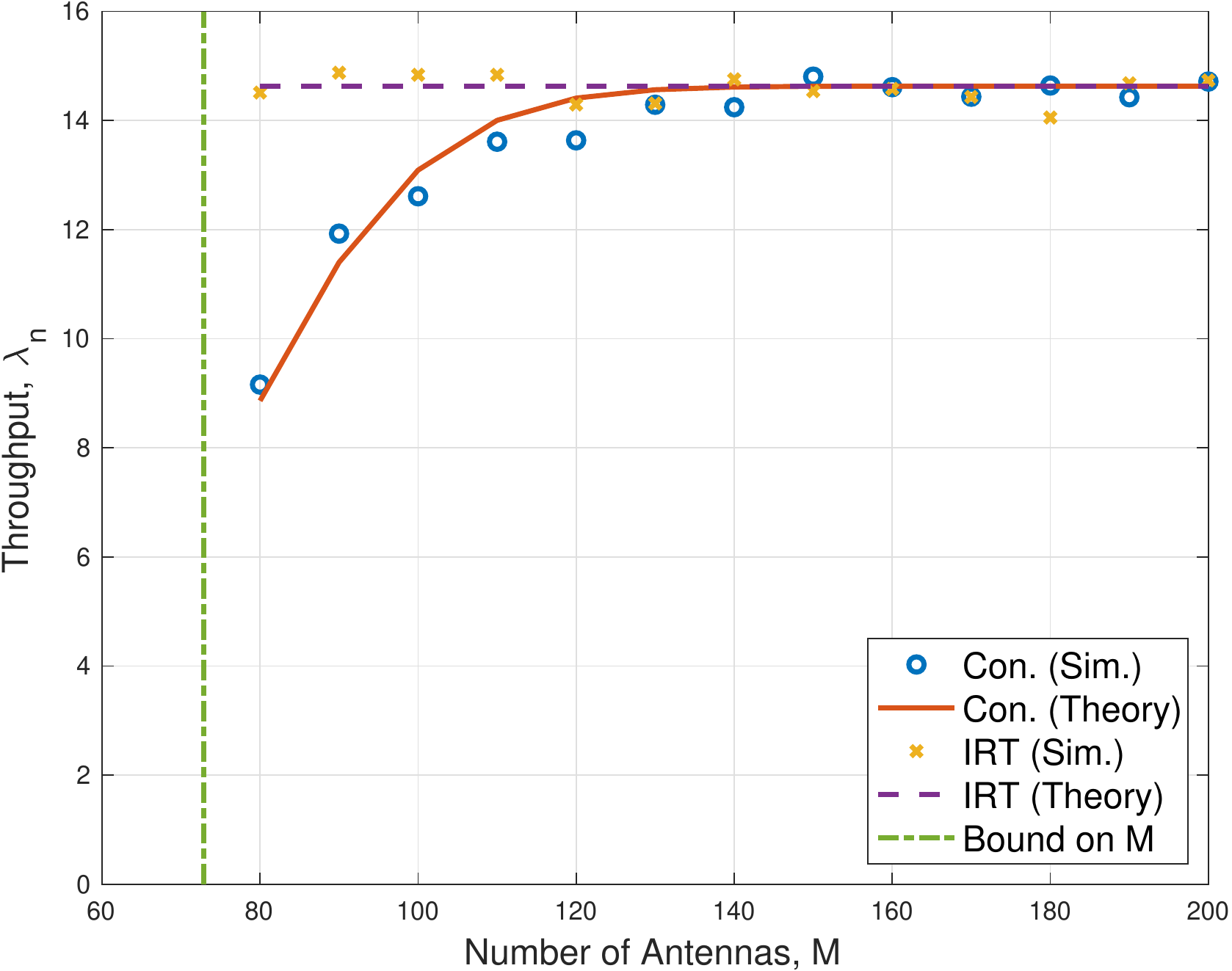} \\
(a) \\
\includegraphics[width=\figwidth]{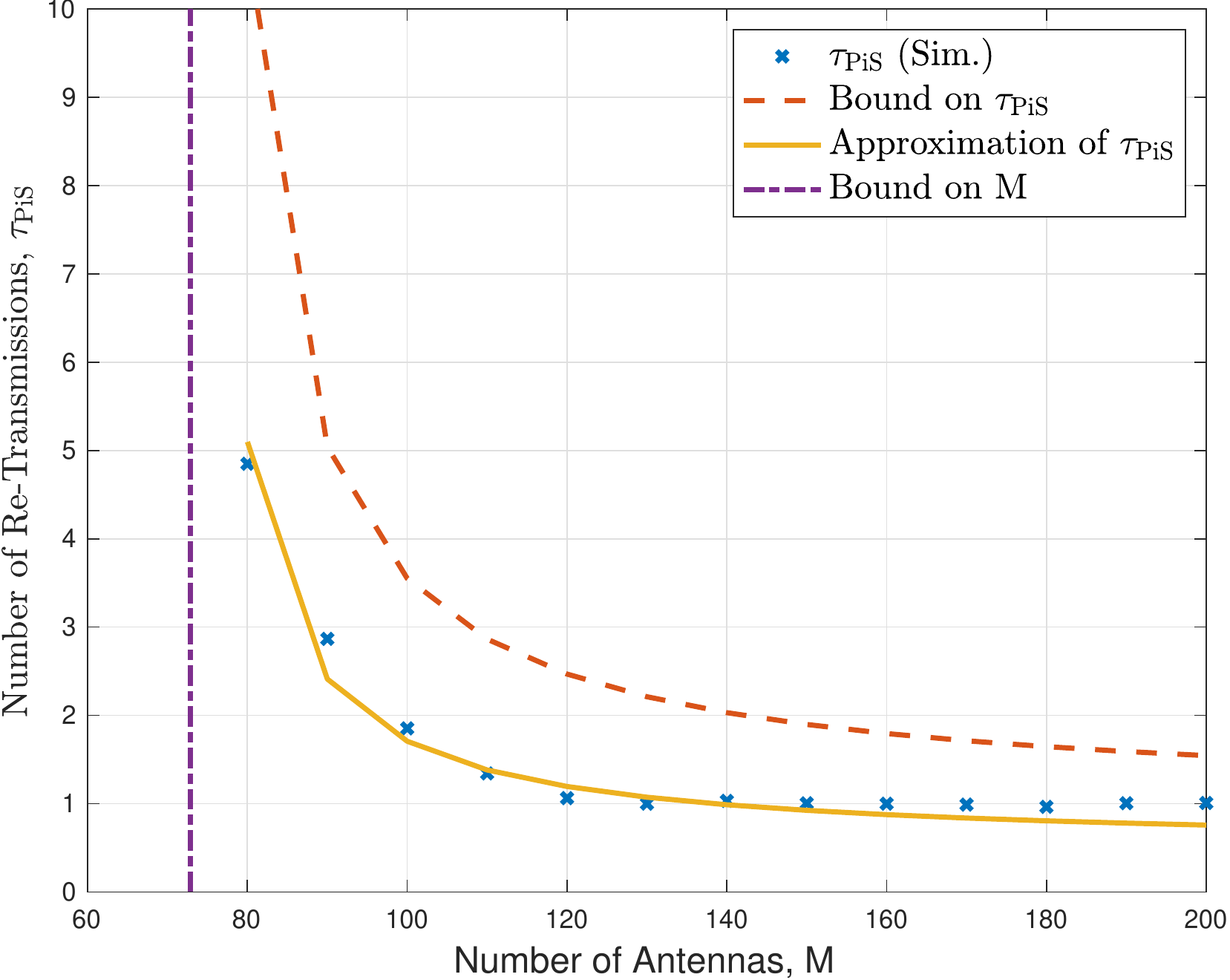} \\
(b) \\
\end{center}
\caption{Performance of the conventional
and IRT systems for different values of 
the number of antennas, $M$,
when $\lambda = 20$, $L = 64$, and $\gamma = \Gamma = 6$ dB:
(a) throughput versus $M$; (b) $\tau_{\rm PiS}$ versus $M$.}
        \label{Fig:plt_thp4}
\end{figure}

To see the impact of the number of
antennas on the performance,
simulations are carried out with different values of $M$
and the results are shown in Fig.~\ref{Fig:plt_thp4}
when $\lambda = 20$, $L = 64$, and $\gamma = \Gamma = 6$ dB.
The bound on $M$ is 
$\bar \lambda \Gamma b_1 = 72.88$
from \eqref{EQ:L2}.
As shown in Fig.~\ref{Fig:plt_thp4} (a),
if $M$ is not too large (e.g., $M < 140$), 
the throughput of the conventional system
is lower than $\bar \lambda = 14.63$,
which is the throughput of stable IRT.
In Fig.~\ref{Fig:plt_thp4} (b),
$\tau_{\rm PiS}$ of IRT is shown,
where we can see that 
$\tau_{\rm PiS}$ becomes finite if 
$M > \bar \lambda \Gamma b_1 = 72.88$.

\begin{figure}[thb]
\begin{center}
\includegraphics[width=\figwidth]{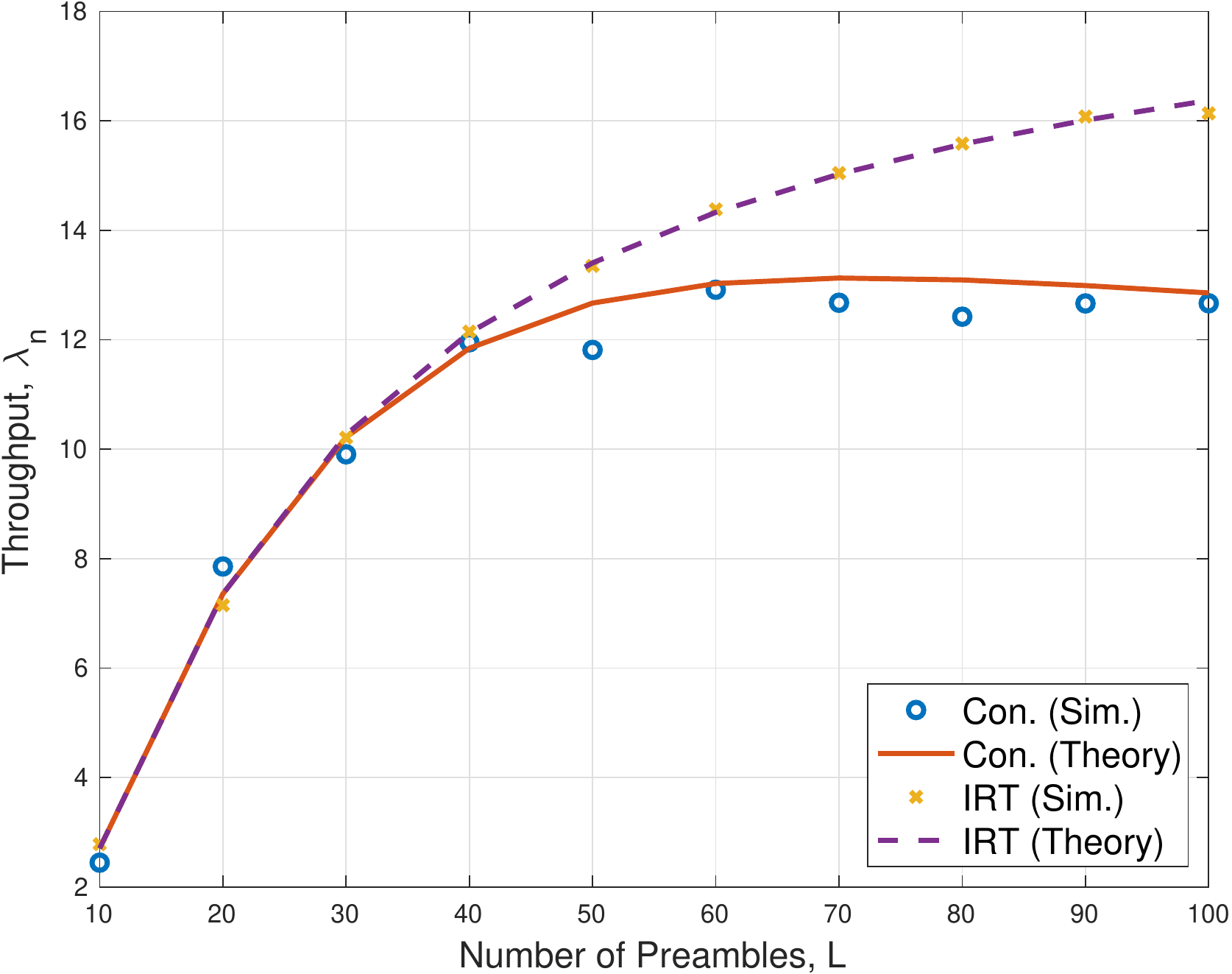} \\
(a) \\
\includegraphics[width=\figwidth]{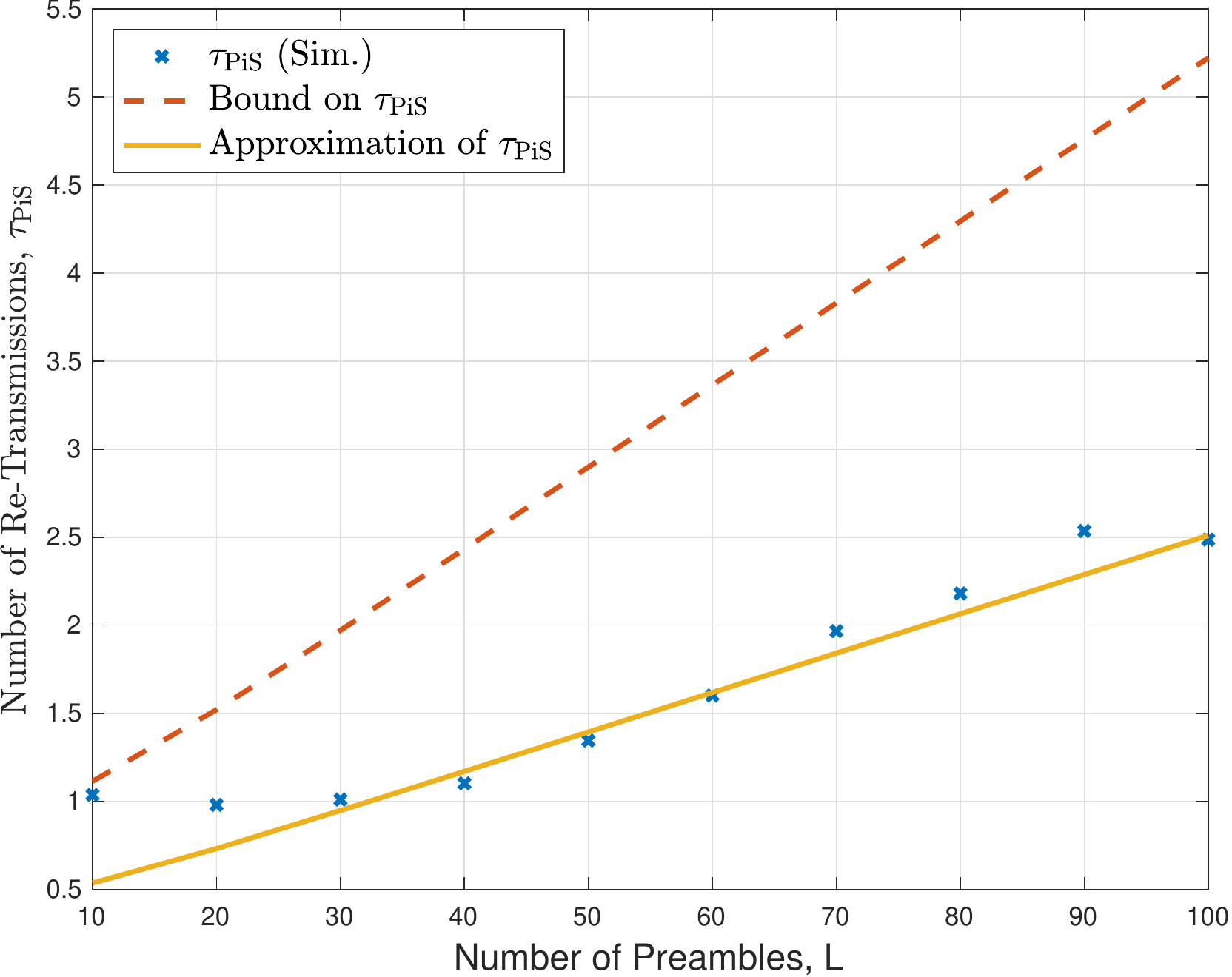} \\
(b) \\
\end{center}
\caption{Performance of the conventional
and IRT systems for different values of 
the number of preambles, $L$,
when $M = 100$, $\lambda = 20$, and $\gamma = \Gamma = 6$ dB:
(a) throughput versus $L$; (b) $\tau_{\rm PiS}$ versus $L$.}
        \label{Fig:plt_thp3}
\end{figure}

In grant-free random access with massive MIMO,
the number of preambles, $L$, is one of key parameters.
Since the probability of PC decreases with $L$,
a large $L$ is expected for a higher throughput.
In Fig.~\ref{Fig:plt_thp3},
the impact of $L$ on the performance is shown with
$M = 100$, $\lambda = 20$, and $\gamma = \Gamma = 6$ dB.
In this case,
since
$$
\bar \lambda \Gamma b_1 \le \lambda \Gamma b_1 = 99.62 < M = 100,
$$
i.e., \eqref{EQ:S2} holds,
the IRT system becomes stable for any value of $L$.
Thus, the throughput of IRT,
$\bar \lambda = \lambda e^{-\frac{\lambda}{L}}$,
increases with $L$ as shown in
Fig.~\ref{Fig:plt_thp3} (a).
On the other hand, in the conventional system,
the increase of $L$ results in the increase of PC-free devices
that leads to the decrease of SINR and 
increase of the probability of unsuccessful decoding.
As a result, the throughput does not increase with $L$ once $L$
is sufficiently large as shown in Fig.~\ref{Fig:plt_thp3} (a).
In the IRT system, 
as mentioned earlier, since the increase of $L$ results
in the increase of $\bar K(t)$, 
$\tau_{\rm PiS}$ can increase with $L$
as shown 
in Fig.~\ref{Fig:plt_thp3} (b).
Thus, there is a trade-off between $\tau_{\rm PiS}$
(or delay) and throughput in terms of $L$
in the IRT system.

\section{Concluding Remarks}	\label{S:Con}

In this paper, in order to improve the
throughput of grant-free random
access with massive MIMO,
we proposed an approach that is based on
immediate re-transmissions for RTD gain.
In particular, for a PC-free active device,
since its channel vector 
is already estimated, no preamble transmissions
are required in immediate re-transmissions. As a result,
without increasing the probability of PC 
in subsequent slots,
the SINR could be increased by immediate re-transmissions for
successful decoding that leads to improved throughput.

The proposed system 
with a finite number of antennas has been analyzed using certain
approximations of SINR with RTD and the resulting analysis provided
key conditions 
in terms of the number of antennas, $M$, and the threshold SINR, $\Gamma$,
to avoid unstable systems 
where the number of re-transmissions can be unbounded.
In addition, the throughput of the proposed system was obtained,
which is shown to be higher than that of the conventional system.
Since the analysis was based on 
approximations of SINR, simulations have been carried out and it was
shown that analysis results reasonably match
simulation results. 

In this paper, we mainly focused on the 
throughput of the proposed system without considering
any re-transmission strategies for the packets of active devices
with PC (i.e., backlogged packets). 
Optimal re-transmission strategies 
for backlogged packets with delay constraints 
and their delay analysis
might be further research topics.
In addition, 
the proposed approach can be generalized with
non-orthogonal preambles for a higher throughput,
which is another further research topic.

\bibliographystyle{ieeetr}
\bibliography{mtc}

\end{document}